\documentclass{PoS}
\usepackage{amsmath}
\newcommand{\veritas}{\textit{VERITAS}}
\newcommand{\magic}{\textit{MAGIC}}
\newcommand{\angstrom}{\text{\normalfont\AA}}

\title{Very-high-energy blazars: updates from \veritas\ observations and multi-wavelength campaigns}
\author{\speaker{Matteo Cerruti} for the VERITAS Collaboration\\
     Harvard-Smithsonian Center for Astrophysics, 60 Garden Street, Cambridge, MA, 02138, USA
     E-mail:  \email{matteo.cerruti@cfa.harvard.edu}}
\abstract{VERITAS is an array of four 12-m atmospheric Cherenkov telescopes, designed to observe the very-high-energy (VHE; E $\geq$100 GeV) sky. Since 2007, it has detected more than 20 extra-galactic sources, the majority of which are active-galactic-nuclei of the blazar class.
In this paper we present a selection of the most recent results from the VERITAS blazar observing program, in particular in the context of broad-band multi-wavelength campaigns. Four results are highlighted: the detection of the flat-spectrum-radio-quasar \textit{PKS\ 1222+216} (\textit{4C +21.35}) during March 2014; the $\gamma$-ray flare from the BL Lac object \textit{1ES\ 1727+502}, observed under bright-moonlight conditions during May 2013; the bright $\gamma$-ray flare from the BL Lac object \textit{1ES\ 1011+496} during February 2014; and the long-term campaign on the BL Lac object \textit{PKS~1424+240}, currently the farthest (z $\geq 0.60$), persistent VHE emitter.}
\ShortTitle{Very-high-energy blazars: updates from \veritas\ observations and multi-wavelength campaigns}

\FullConference{10th INTEGRAL Workshop: "A Synergistic View of the High Energy Sky" - Integral 2014,\\
		15-19 September 2014\\
		Annapolis, MD, USA }

\begin{document}

\section{Introduction}

A non-thermal continuum from radio to $\gamma$-rays characterizes blazar emission. This, together with rapid variability (with time-scales as short as minutes), implies that blazar physics is accessible only via multi-wavelength (MWL) campaigns, in order to build simultaneous spectral-energy-distributions (SEDs) of these sources. Blazar monitoring programs are in place at every wavelength, from radio to very-high-energies (VHE; E$>$100 GeV).\\

The observational properties of blazars (variability, polarization, broad-band non-thermal continuum) are explained in the framework of the unified model of active galactic nuclei (AGN) considering that blazars are  radio-loud AGN whose relativistic jet is closely aligned with the line of sight (see \cite{Urry}).\\

The blazar class is composed of flat-spectrum-radio-quasars (FSRQs) and BL Lac objects: while the former present emission lines in their optical/UV spectrum, the latter are characterized by weaker lines (equivalent width $<$ 5 \angstrom) or even a featureless non-thermal emission. The two sub-classes are characterized as well by different luminosity and redshift distributions, with FSRQs being in general brighter and located at higher redshifts than BL Lac objects. 
Blazar SEDs are composed of two different components, peaking in infrared-to-X-rays and $\gamma$-rays, respectively. While FSRQs are in general characterized by the frequency of the first SED peak in infrared, BL Lac objects show a variety of peak frequencies, and are further classified (see \cite{Padovani}) into low-frequency-peaked BL Lac objects (LBLs, with peak in infrared), intermediate-frequency-peaked BL Lac objects (IBLs, with peak in optical) and high-frequency-peaked BL Lac objects (HBLs, with peak in UV and above).\\

The population of blazars detected at VHE is not homogeneous: among the 54 blazars detected so far\footnote{For an updated list of VHE blazars see http://tevcat.uchicago.edu}, HBLs constitute 80\%, and only four FSRQs have been detected. In this paper we will present some recent results on VHE blazars from the \veritas\ telescope array. In Section \ref{sec2} we will introduce the \veritas\ telescopes, and in Section \ref{sec3} we will discuss four objects in details: the FSRQ \textit{PKS 1222+216} and the HBLs \textit{1ES 1727+502}, \textit{1ES 1011+496} and \textit{PKS 1424+240}.

\section{The \veritas\ Telescope Array}
\label{sec2}

The \veritas\ telescope array is a system of four Cherenkov telescopes, located at the Fred Lawrence Whipple Observatory (FLWO) in Southern Arizona. The telescopes are 12-m diameter reflectors, arranged in a square with sides of $\sim 100$ m. The instrument field of view is $3.5^\circ$, with an angular resolution of $\sim 0.1^\circ$. \veritas\ is sensitive to VHE $\gamma$-rays with an energy of 100 GeV to $\sim$30 TeV, and the energy resolution is $15-20\%$. For further details on the instrument see e.g. \cite{Holder}.\\

In operation since 2007, \veritas\ has undergone two major upgrades: in 2009 one of the telescopes was relocated, obtaining a symmetric array; in 2012 the camera photo-multipliers (PMTs) were replaced, lowering the energy threshold, and increasing the overall sensitivity. \\

Observations by Cherenkov telescopes are limited by the sensitive PMTs, which cannot safely operate under bright moonlight: \veritas\ standard observations are performed only with a Moon illumination lower than $35\%$. Since 2012, the \veritas\ collaboration has worked with a new program in order to extend the observatory duty cycle, and perform observations under bright-moonlight conditions. This is achieved by either applying a reduced high-voltage (RHV) to the camera PMTs, or by installing UV-transparent filters. The extension of the duty cycle is particularly important for variable sources, such as blazars. The first scientific result of this new observing strategy is the detection of $\gamma$-ray flaring activity from the HBL  \textit{1ES 1727+502}, and will be discussed in the next section.\\

\section{Selected \veritas\ results}
\label{sec3}

\subsection{PKS 1222+216}

Also known as \textit{4C +21.35}, the FSRQ\,  \textit{PKS 1222+216} (z=0.432) is one of the only four FSRQs ever detected at VHE by Cherenkov telescopes. It was discovered by the \magic\ telescopes during an extremely bright $\gamma$-ray flare ($\sim1$ Crab, in January 2010), showing emission up to 400 GeV and a variability time scale of $\sim 10$ minutes (see \cite{1222Magic}). \\

During February 2014, the source showed enhanced activity in optical (\cite{1222ATelop}) and in the MeV-GeV energy range (as detected by Fermi-LAT). The \veritas\ telescopes observed \textit{PKS 1222+216} starting 11 days after the Fermi-LAT flare, and continued monitoring the source for the following ten nights (February 26, 2014 - March 10, 2014) for a total exposure of $\sim 5$ hours.\\

\begin{figure*}
	   \centering
		\includegraphics[width=212pt,height=180pt]{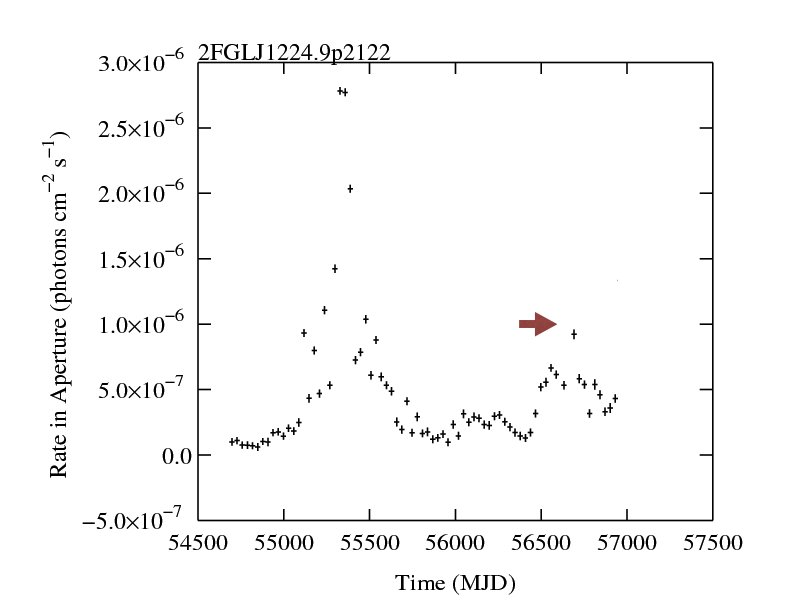}
		\includegraphics[width=212pt,height=180pt]{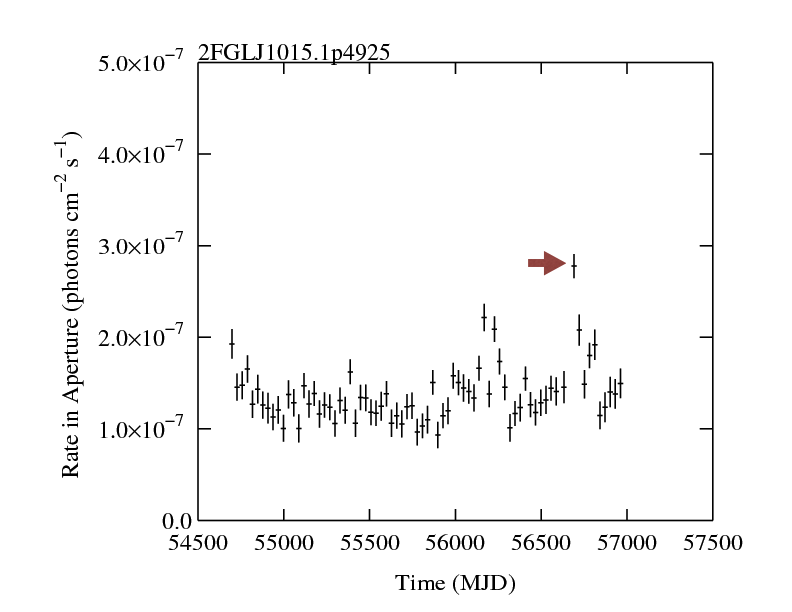}
	  \caption{Results from the automatic light-curve aperture-photometry pipeline provided by the Fermi-LAT team (http://fermi.gsfc.nasa.gov/ssc/data/access/lat/2yr\_catalog/ap\_lcs.php). \textit{Left:}  \textit{PKS 1222+216}; \textit{Right:}  \textit{1ES 1011+502}. The red arrow indicates the $\gamma$-ray flares that are discussed in the text. \label{figone}}
   \end{figure*}

The source is detected at $\sim 6$ standard deviations, and the flux level is preliminary estimated as $3\%$ of the Crab Nebula above 100 GeV (see \cite{1222ATel}). Interestingly, there is no indication of variability during the 10 days of \veritas\ observations, contrary to the first \magic\ detection. In Fig. \ref{figone}  we show the public Fermi-LAT light-curve, highlighting the bright flare, and the period of \veritas\ observations. Simultaneous monitoring in X-rays and UV has been provided by the Swift satellite, showing that the source was active in X-rays, as well. The data-analysis of the MWL campaign is currently in preparation.\\

\subsection{1ES 1727+502}
\begin{figure*}
	   \centering
		\includegraphics[width=220pt,height=280pt]{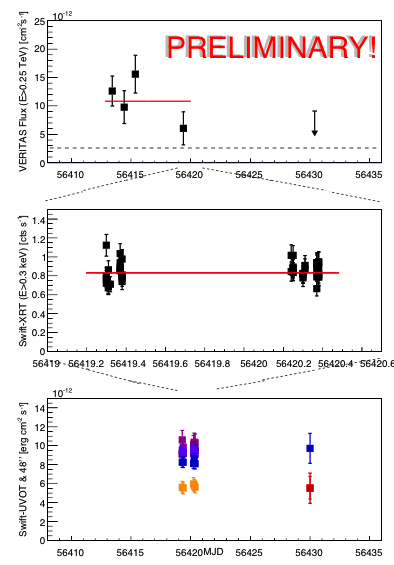}
		\includegraphics[width=340pt,height=230pt]{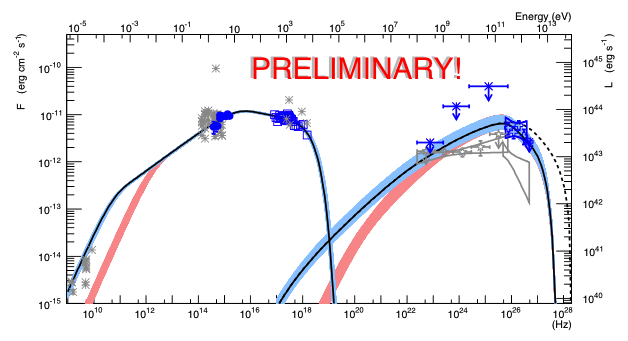}
	  \caption{\textit{Top:} Preliminary multi-wavelength light-curve of \textit{1ES 1727+502}. From top to bottom, \veritas,  Swif-XRT and Swift-UVOT / 48". \textit{Bottom:} SED of \textit{1ES 1727+502} during the 2013 flaring activity; the SED is modeled with a one-zone synchrotron-self-Compton model \cite{SSCconstraints}.\label{fig1727}}
   \end{figure*}

Located at a redshift $z=0.055$, the HBL 1ES 1727+502 was first detected at VHE by \magic\ (\cite{1727Magic}), with an integral flux above $150$ GeV equal to $(2.1 \pm 0.4)\%$ of the Crab Nebula, and with a spectral index $\Gamma = 2.7 \pm 0.5$. The source showed increased $\gamma$-ray activity during May 1, 2013, detected by \veritas\ in the RHV configuration. Additional observations were triggered during the following days, together with the Swift satellite. \\
The MWL campaign included Swift and the optical 48" telescope at the FLWO (although observations from the latter were not simultaneous with \veritas).  The VHE flare was not detected by Fermi-LAT, and only an upper limit could be computed for the MeV-GeV part of the SED. No significant variability was detected in any energy band. The \veritas\ light-curve is consistent with constant emission from May 1 to May 8, 2013. The source was not detected by \veritas\ on May 18, 2013, indicating that the high-flux state may have ended at some point after May 8. The best-fit of the VHE emission results in an integral flux above 250 GeV of ($1.1 \pm 0.2$)$\times\ 10^{-11}$ cm$^{-2}$ s$^{-1}$ , with an index $\Gamma=2.1 \pm 0.3$. This corresponds to roughly five times the flux measured at the time of the archival \magic\ detection (above 250 GeV).  In Fig. \ref{fig1727} the SED of 1ES 1727+502 is shown: it represents the first quasi-simultaneous SED of this blazar during $\gamma$-ray flaring activity. The stationary particle population is fully consistent with leptons injected with a power-law distribution with index $\alpha_1=2.2$, simply cooled by synchrotron radiation. The paper showing the details of the MWL campaign is currently in preparation, and will represent the very first \veritas\ paper using bright-moonlight data.

\subsection{1ES 1011+496}

The HBL \textit{1ES 1011+496} was discovered in VHE by \magic: the flux above 200 GeV was estimated as $7\%$ of the Crab Nebula. At a redshift of 0.212, at the epoch of the discovery, it represented the most distant VHE blazar (see \cite{1011Magic}). \\
During February 2014, \veritas\ observed a strong $\gamma$-ray flare. In the following day the source was monitored by both \veritas\ and \magic, showing an integral flux level between $25$ and $75\%$ of the Crab Nebula (see \cite{1011ATel}), i.e. up to a factor of ten brighter than its baseline flux.\\
The VHE flare was associated as well with a GeV flare, as shown by Fermi-LAT data (see Fig. \ref{figone}). The Swift satellite covered the UV and X-ray part of the spectrum, detecting rapid variability. The extremely rich data-set is currently under study, and will be presented in details in an upcoming publication.\\

                   \begin{figure*}
	   \centering
		\includegraphics[width=260pt,height=220pt]{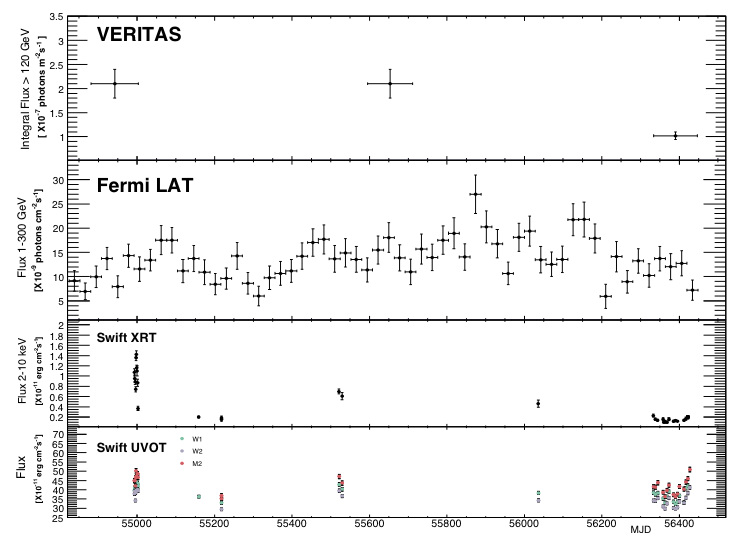}
		\includegraphics[width=320pt,height=260pt]{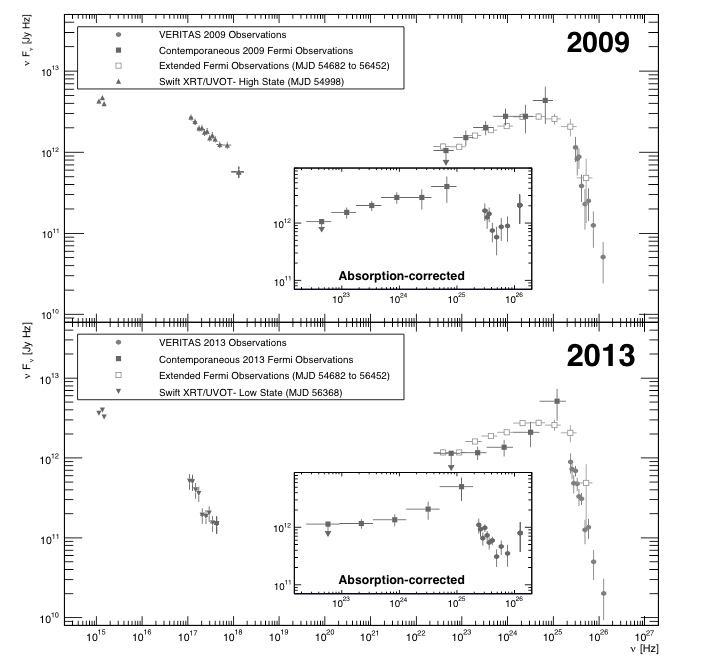}
	  \caption{\textit{Top:} Multi-wavelength light-curve of \textit{PKS 1424+240}, taken from \cite{1424two}. From top to bottom, \veritas, Fermi-LAT, Swif-XRT and Swift-UVOT. \textit{Bottom:} SED of \textit{PKS 1424+240} during 2009 and 2013; the absorption-corrected VHE spectrum is shown in the inset. \label{fig1424}}
   \end{figure*}

\subsection{PKS 1424+240}

Located at a redshift higher than 0.60 (as constrained by the detection of absorption lines due to intergalactic material in the line of sight, see \cite{Furniss1424}), \textit{PKS 1424+240} is currently the most distant, persistent VHE emitter\footnote{The \magic\ collaboration recently reported the detection of VHE emission from the gravitationally lensed blazar \textit{S4 0218+35} (z=0.944) during flaring activity (see \cite{ATel0218}).}. The blazar is a HBL and its VHE emission has been discovered by \veritas\ in 2009 (\cite{1424one}). At that time, the redshift of the source was unknown, and the VHE emission from \textit{PKS 1424+240} did not receive any particular attention.  \\
Following the first redshift estimation, in 2013 \veritas\ started a new deep observing campaign, reaching more than 100 hours of exposure, and allowing spectral reconstruction up to 750 GeV. The details of the three-year campaign, including data from Fermi-LAT and Swift, are discussed in \cite{1424two}. \\
 The high-redshift of this blazar implies that its VHE emission is significantly absorbed by pair-production over the extra-galactic background light (EBL). In Fig. \ref{fig1424} we show as well the intrinsic (i.e. corrected for EBL absorption) source emission. Reference \cite{Furniss1424} suggested that, given the distance of the source, the intrinsic emission may show evidence of a hardening at high-energies. With the new \veritas\ data, the significance of the spectral hardening remains marginal. Further observations are required to improve the statistics and quantify the presence (or not) of such a hardening. If detected without associated variability, it would lend support to theoretical models which suggest that the blazar emission is associated to cosmic-rays triggering cascades along the path from the source to the observer \cite{Essey}; or can indicate new physics such as axion-like particles (see  \cite{Horns}).

\section{Conclusions}

Blazar physics is accessible only via simultaneous MWL observations, involving multiple instruments covering the entire electro-magnetic spectrum. The \veritas\ array is one of the best telescopes to study blazars at VHE. In this paper, we presented four selected, recent results from the \veritas\ blazar observations. All of them are based on MWL campaigns, and permit the study of flux correlations at different energies, and the building of simultaneous broad-band SEDs. In the recent years the focus of \veritas\ shifted from discovery targets to long-term monitoring of known VHE blazars \cite{Benbow}, increasing the possibility of catching blazar flares. Among the results presented here, the flares detected from \textit{1ES 1011+496} and \textit{1ES 1727+502} indeed represent the brightest VHE flares observed so far from these two sources.\\
In this context, the new \veritas\ observing strategy under bright-moonlight, increasing the available observing time, will be particularly useful for blazar studies, extending the monitoring capabilities at VHE.\\

\end{document}